# Remote Home Management:

An alternative for working at home while away

[1]B. I. Ahmad, [2]F. Yakubu, [3]M. A. Bagiwa, [4]U. I. Abdullahi

[1]Department of Mathematics, Ahmadu Bello University, Zaria. Kaduna, Nigeria
[2]Iya Abubakar Computer Center, Ahmadu Bello University, Zaria. Kaduna, Nigeria
[3]Department of Mathematics, Ahmadu Bello University, Zaria. Kaduna, Nigeria
[4]Department of Mathematics, Ahmadu Bello University, Zaria. Kaduna, Nigeria

[1]barroonia@yahoo.co.uk, [2]yakfri@yahoo.com, [3]mstphaminu@yahoo.com, [4]amaall1010@yahoo.com

*Abstract*—Remote home management is one of the developing areas in current technology. In this paper we described how to manage and control home appliances using mobile phone, people can use this system to do things in their home from a far place before they reach home. For instance, user may start his/her room cooler or heater so that before they reach home the condition in the room will be conducive, also appliances like washing machine and cooker can be started and if the time taken for this appliances to perform a task is known that can also be set, so that if the time elapsed the appliance will automatically switch off itself. To control an appliance the user sends a command in form of SMS from his/her mobile phone to a computer which is connected to the appliance, once the message is received the computer will send the command to a microcontroller for controlling the appliance appropriately.

## I. INTRODUCTION

Mobile phones are getting more advanced which allow people to develop applications that run on them. Currently mobile phones are gradually replacing PCs because of their ability to do almost all whatever computers can do.

Remote management and control of devices is one of the areas where an application can be developed to makes our life easier. Different approaches can be followed to develop remote management or control systems, some use DTMF (Dual Tone Multi Frequency) technology which involves using mobile phones tone to perform an action, while some use SMS technology to send the command for a particular action, some also use GPRS (General Packet Radio Service) technology to directly interface mobile phone and the computer.

## II. RELATED WORK

A lot of research work has been carried out in the area of wireless and mobile computing, as a result of which the area is becoming very wide. Even though, there are still some challenges in the area like using database systems for mobile [6], security, streaming of data, etc...

A system of controlling devices using DTMF was developed [1], in which a user has to call a phone connected to the system via head set, then after the call is answered (auto answer) the user would enter two/three digit code. When the user presses these codes a specific operation is performed by microcontroller, which can be ON or OFF a specific device, a code that indicate completion of a function would be sent back to the user's mobile phone. This system also provides security by means of alarming the user when there is an intrusion.

In another paper [2], home/office remote controller equipped with power controllers, an alarm system, a voice memory and backup battery unit was developed. When the user switch ON alarm system, it monitors fire and PIR (Passive Infrared) detectors, the system allow the user to use up to 8 detectors, whenever there is fire outbreak the system can call up to 5 numbers including fire brigade, police, and the owner of the premises and provide the pre-recorded voice message. There is a unit that supplies power to the system in case of power failure.

We can have an integrated system for remote controlling devices like GSM based Remote Monitoring and Control System[3], which is used for controlling, monitoring and accessing the performance of remotely situated device parameters such as Temperature, Pressure, and Humidity etc. Whenever the value crosses a certain limit, an SMS will be sent to the user as a notification.

Joude et al developed a system called Home Management System using a GPRS-Enabled Device [4]; the system is used to control household devices like fan, light, and water heater. The main part of the system are divided into three: the client which is a mobile phone with a software installed for selecting the device and the operation intended to perform on the device, server which is a computer with database for storing the status of the device, and a hardware that control the devices connected to the server via serial communication. Apart from handling the devices using General Packet Radio Service (GPRS), the system sends notification to the user if the device status changes.

Some systems were developed with the option of either using mobile phone or a computer to control devices like Home Network System [5]. This is an integrated system that uses GSM mobile or a computer for monitoring and controlling of devices, the system comprised of HNS gateway for receiving and managing the messages and three subsystems. The subsystems are: home appliance like fan, bulb, etc; security which uses charged-coupled device (CCD) camera for surveillance and messaging which uses email as a sample application.

### III. SYSTEM STRUCTURE

This system is structured as shown in Figure 1. The system consist of two mobile phones, one to be used by the user while the other for receiving the message sent by the user and for sending reply back to the user.

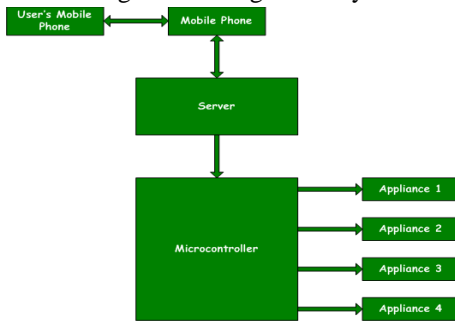

Fig. 1: System Block Diagram.

There is a computer (server) as part of the system, which stores the message received and uses the content of the message to know which appliance the user wants to control. Controlling the appliances is achieved using a circuit that sends the command to the appliances. The computer is connected the circuit via parallel connection, while the mobile phone is connected to it via serial connection.

### IV. SYSTEM COMPONENTS

The main components of the system are described below.

#### A. User's Mobile Phone

The first part of the system is the user's mobile phone, which is a phone that supports Java. We need Java enabled phone because we have an application developed using Java 2 Micro Edition (J2ME) technology for sending the message to control the appliances. We can use the normal Short Message Service (SMS) provided by the manufacturers of mobile phone, but in that case people can see the mobile number of the phone that will receive the message which may increase the risk of unauthorized access to the system; also when sending the message the user can make mistake in typing the number. To solve the problems above we decided to have an application that will run on the users mobile phone with the phone number of the receiving phone embedded. The application also expects the user to supply right login credentials for authenticating the user before having access to it.

#### B. Server

The server in this system is a computer connected with a mobile phone for receiving the message from the user. The connection is done via RS232, when a message is received it will first enter the mobile phone then immediately an SMS Gateway Application called Gammu download the message into the computer and store it in a database. The content of the message is checked upon received to know what appliance the user wants to control and what operation (on/off) to be performed.

This is the main component of the system where we have the core functionality; we have a program that retrieves the message received from the database and use the message content to decide on which decision to take. The decisions include, the appliance to act on, the operation to perform on that appliance (ON, OFF, or ON for some times). When the decision is taken, the server will then sends data to one of the cables connected to the microcontroller that will either switch ON or OFF the appliance. Since any time the user can send a message, the server needs to always be ON.

#### C. Microcontroller

The microcontroller consists of diodes, resistors, transistors and relay for proper control of the appliances. The application can only be used to control one appliance at a time, but you can have one appliance started for sometimes and before it is stopped you can control another. The circuit diagram is depicted in figure 2.

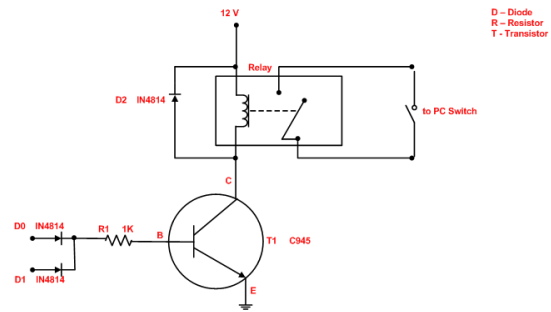

Fig. 2: Circuit Diagram.

#### D. Devices

The devices are the appliances that the user wants to control; user can control up to four appliances using this system. The appliance can be a washing machine, cooler, boiler, cooker, etc. Any of these devices can be started while the user is away and stopped as desired. Devices that may cause problem if not stopped on time like cooker, we make them in such a way that once they are started there is a time limit for them to be stopped. But user can start appliances like Air Conditioner and stop them any time they wish.

## V. PROCESS FLOW

The first step in using this system is to write and send the command in form of SMS to the mobile phone connected to the server, when the message is received it will now be stored in Database (MySQL). The content of the message is examined to know what operation the user wants to perform; the whole process is illustrated in figure 3.

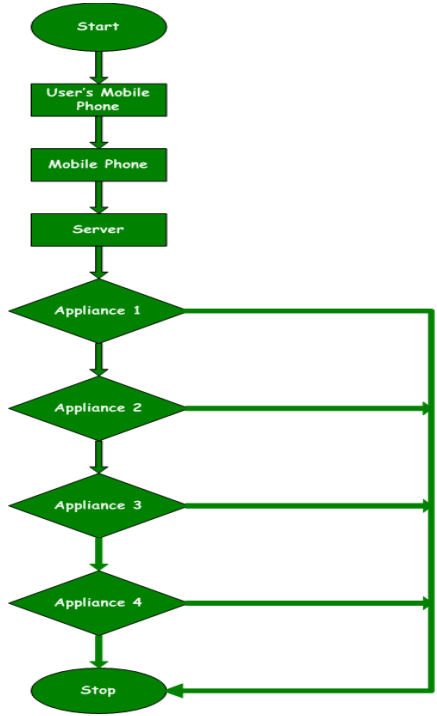

Fig. 3 Process Flow.

In this part we have a series of decision statements for selecting the appliance, type of operation, etc.

The message body must have two or three fields, two fields message means a user wants to perform either ON or OFF operation as shown in figure 4, while three fields message as depicted in figure 5 is for starting an appliance and stop after some times.

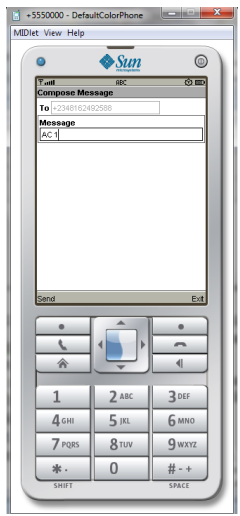

Fig. 4 Sample message (ON).

The first field represents the appliance name, the middle is a number that represent the type of operation; one is ON while zero is OFF, while the last field represents the time an appliance will take to be ON. Each time a message is sent, it has to be stored in the database and the message has to be broken down into fields, then the program on the server take the values for those fields and send a command specific to the type of operation to microcontroller which will then act on the appliance. For instance, the diagram in figure 4 is a message that will start an AC, while that of figure 5 is for starting a cooker for 1800 seconds (30 min).

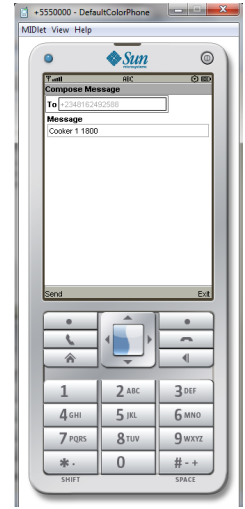

Fig. 5 Sample Message (ON for sometimes).

## VI. CONCLUSION

This paper described how we can develop a remote management and control system for home appliances, the system not only allows the user to start or stop appliances but can also specify how long did he/she wants an appliance to be ON. With this system user can control up to four devices.

There are limitations for this system which brings about the need for further improvement, some of which include:
- Controlling multiple appliances concurrently.
- Sending a confirmation message back to the user.
- Be able to know the status of the appliances.
- Add DTMF technology as an alternative to SMS.

.

AUTHORS PROFILE

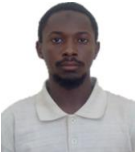
Barroon Isma'eel Ahmad received the B.Sc. Degree from Usmanu Danfodiyo University Sokoto Nigeria, in 2006. He joined Department of Mathematics, Ahmadu Bello University, Zaria as Graduate Assistant in 2008, and currently he is a Postgraduate Student in the same Department. His interests are Networking and Mobile Application Development.

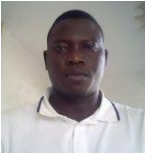
Friday Yakubu received the B.Tech. Degree from Abubakar Tafawa Balewa, University, Bauchi, Nigeria, in 2003, he obtained Masters of Information Management from A.B.U Zaria in 2008. He joined Ahmadu Bello University, Zaria in 2006. His interests are Networking and Web Based Programming.

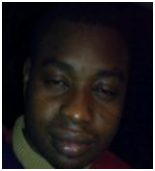
Mustapha Aminu Bagiwa is an M.Sc. Student of Computer Science in Ahmadu Bello University, Zaria. Nigeria. He received his B.Sc. Degree in computer science from Usmanu Danfodiyo University Sokoto, Nigeria in 2006. He is currently working as a Graduate Assistant in the Department of Mathematics, Ahmadu Bello University, Zaria. Nigeria.

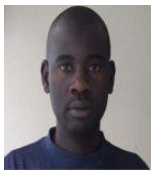
Umar Isyaku Abdullahi obtained his first degree in mathematics from Ahmadu Bello University Zaria, Nigeria in 2006. He attended Robert Gordon University United Kingdom where he obtained MSc in computing: Software Technology in 2009. His research interests are: Machine learning, Mobile application, Application of Statistical Computing in Sampling Survey, and Statistical Modeling.